\documentstyle[11pt,aaspp4,psfig]{article}
\renewcommand{\arraystretch}{1.0}

\lefthead{Daflon \& Cunha}

\righthead{High $v \sin i$ B stars}

\begin{document}

\title{Chemical Abundances of OB Stars with High Projected Rotational
Velocities}

\vskip 1truecm
\author{Simone Daflon}
\affil{Observat\'orio Nacional, Rua General Jos\'e Cristino 77 \\
CEP 20921-400, Rio de Janeiro  BRAZIL \\
daflon@maxwell.on.br \\} 
\author{Katia Cunha}
\affil {Observat\'orio Nacional, Rua General Jos\'e Cristino 77 \\
CEP 20921-400, Rio de Janeiro  BRAZIL \\
katia@gauss.on.br \\} 
\author{Keith Butler}
\affil{Institut f\"ur Astronomie und Astrophysik der Universit\"at M\"unchen, 
Scheinerstrasse 1, D-81679 M\"unchen, GERMANY \\
butler@usm.uni-muenchen.de \\}
\author{Verne V. Smith}
\affil{Department of Physics, University of Texas at El Paso \\
 El Paso, TX 79968-0515 USA \\
verne@barium.physics.utep.edu \\} 
\clearpage

\begin{abstract}

Elemental abundances of carbon, nitrogen, oxygen, magnesium,
aluminum, and silicon are presented for a sample of twelve rapidly 
rotating OB star  ($v \sin i > 60 {\rm km s}^{-1}$) members of the
Cep OB2, Cyg OB3 and Cyg OB7 associations.  The abundances 
are derived from spectrum synthesis, using both LTE and non-LTE
calculations. As found in almost all previous studies
of OB stars, the average abundances are slightly below solar, by about
0.1 to 0.3 dex. In the case of oxygen, even with the recently derived low
solar abundances the OB stars are closer to, but still below, the
solar value. Results for the 9 Cep OB2 members in this sample can
be combined with results published previously for 8 
Cep OB2 stars with low projected rotational velocities to yield the most 
complete set of abundances, to date, for
this particular association.  These abundances provide a clear picture 
of both the general chemical and individual stellar evolution
that has occurred within this association. By placing the Cep OB2 stars
studied in an HR diagram we identify
the presence of two distinct age subgroups, with both subgroups
having quite uniform chemical abundances.  Two stars are found in the older
subgroup that show significant N/O overabundances, with both stars being
two of the most massive, the most evolved, and most rapidly rotating
of the members studied in Cep OB2.  These characteristics of increased
N abundances being tied to high mass, rapid rotation, and an evolved phase
are those predicted from models of rotating stars which undergo
rotationally driven mixing.

\end{abstract}

\clearpage

\section{Introduction}

The determination of the chemical abundance distributions in stars often
requires the identification and isolation of a suitable set
of absorption lines whose individual line strengths, or
equivalent widths, are well-defined and can be measured. These
equivalent widths become the primary datasets upon which
the abundance analyses of the stars are based. For early-type stars that
have small values of $v \sin i$, defining a set of spectral lines whose equivalent
widths can be measured accurately is usually a straightforward
task, because their spectra are relatively free of blending lines. 
As rotational velocity increases and spectral line-widths
broaden, however, a point is reached beyond which it becomes 
impossible to find adequate sets of unblended spectral lines
on which to anchor a detailed abundance analysis. 
The distribution of projected rotational velocity, 
$v \sin i$, as a function of spectral type increases from F-G stars, where 
$v \sin i$ almost vanishes, to OB stars that typically rotate rapidly. 
Wolff {\it et al.} (1982)
studied the distribution of rotational velocities in early-type stars and 
observed that, among hot stars, early B-type stars show the 
lowest rotational velocities;
$ <v \sin i> = 110 {\rm km s}^{-1}$. However, even at the low end,
these values of $v \sin i$  are high enough to
blend the absorption lines to such an extent that prevents
line identification and individual equivalent-width measurements. In this
case, spectrum synthesis can be used to model the blended line
profiles and extract elemental abundances.

Our previous studies of chemical abundances in OB stars 
- Daflon, Cunha \& Becker (1999) and Daflon {\it et al.} (2001),  
hereafter Papers I and II, respectively - analyzed only 
those observed targets with sharp lines 
($v \sin i < 60 {\rm km s}^{-1}$) as the method adopted
to derive elemental abundances was based on  
equivalent-width  measurements.
Due to this restriction, the final list of OB stars analyzed in these previous
papers consisted of about 25\% of the total number of observed targets
for this project. However,  
many of the observed stars with relatively high $v \sin i$ still have 
conspicuous features in their spectra that can be compared with broadened 
synthetic line-profiles
and thus provide information
about their chemical compositions. This paper focuses on a subset of
the observed sample OB stars having $v \sin i$ between roughly 60 and 
140 ${\rm km s}^{-1}$ with
the intent being to derive stellar
parameters and chemical abundances. 
Four stars are included from Paper I
(HD 205948, HD 207951, HD 209339 and HD 239729
of the Cep OB2 association) and 3 stars from Paper II 
(HD 227696 and HD 228199 of Cyg OB3 and HD 202347 of Cyg OB7) for which
stellar parameters were already derived but an abundance analysis was not
done. In addition to these 7 stars, 5 other 
stars belonging to the Cep OB2 association are analyzed here. 
The addition of this set of
rapidly rotating stars to those with lower projected rotational
velocities analyzed
already, enlarges our sample of observations for   
the current study of chemical abundances of OB stars in the Galactic disk as much as possible.
In particular, this completes the sample for the study of the abundance
distribution and investigation of possible inhomogeneities 
in the chemical composition of the gas that formed the main-sequence members 
(with spectral types between O9 and B3 and
for which an abundance analysis can be done) of the Cep OB2 association.

In addition to their usefulness as probes of the general abundance
distributions found in the Galactic disk and in OB associations,
rapidly rotating B-stars can have their surface abundances altered by
rotationally induced mixing.  In a recent study, Heger \& Langer (2000)
have presented surface abundance predictions for elements sensitive to
mixing, such as boron, carbon, or nitrogen, in a set of stellar models
of differing mass, rotational velocity, and age.  The members of a
typical OB association are all formed in the same place, but at slightly
different times (with age differences of $\sim$10$^{5}$--10$^{7}$ yrs),
with differing masses and rotation velocities.  Mass, age, and rotation
are all variables that play a role in the degree of stellar
mixing expected.  In this paper certain elements sensitive to mixing, such as, 
C, N, and O, are analyzed in Cep OB2 and the results can be used 
to test the rotating stellar models.

\section{ Observations and Stellar Parameters}

The observations and data reduction are described in Papers I and II, however,
a brief summary of the instruments follows. 
The data consist of spectra with a resolution of R$\sim$ 60,000, high signal-to-noise 
(S/N $\ge$150), covering the spectral range from 
4225--5285\AA\ and obtained with the 2.1m Otto M. Struve telescope 
plus Cassegrain
Sandiford echelle spectrometer at the McDonald Observatory of the 
University of Texas. Additional spectra were obtained with the 2.7m 
Harlan J. Smith telescope, also at McDonald Observatory,
and Coud\'e spectrometer having R=12,000 and centered on the region of 
$H\gamma$ (4340\AA). 

In Paper I, an effective temperature
scale for OB stars was derived using the reddening-free Q
parameter from UBV photometry. The T$_{\rm eff}$-Q calibration, when combined
with an analysis of the H$\gamma$ line-profiles, has been found to be adequate
for the derivation of the fundamental stellar parameters, T$_{\rm eff}$ and 
log g, needed to conduct an abundance analysis for the main-sequence OB stars.
The adopted stellar parameters for the studied stars 
are collected in Table 1. Most of these are from
Papers I and II and in this study we add the determinations for 5 
rapidly rotating stellar members
of the Cep OB2 association: HD235618, HD239681, HD239710, HD239745
and HD239748. All of the additional stellar parameters presented here
were derived following the same approach adopted in Paper I.
(See that paper for a more detailed discussion of the 
method). LTE plane-parallel model atmospheres were then calculated
with the ATLAS9 code (Kurucz, 1991) for solar metallicity and a constant
microturbulence velocity of 2kms$^{-1}$.
These model atmospheres were adopted for both the LTE and non-LTE 
abundance analyses.

\section{Analysis}

As a matter of consistency with our previous studies, we tried as
much as possible to use the same sets of transitions of C, N, O, Si, Mg and Al
selected in Papers I and II.  However,
because of the high rotational velocity of the target stars this was not
always possible. In particular, the weak lines
of C II, S III and Fe III become too shallow in stars with
relatively high $v \sin i$.  As a result of this, sulfur and 
iron abundances were not derived in this study.
Because of the lack of C II lines, an attempt was made to derive 
carbon abundances using C III lines, as we are interested in complementing 
the sample for carbon as much as possible.
The only possibility was to use the 3 strong C III lines in the region centered 
on 4650\AA, which were not used in our previous papers 
because these strong lines are
very close to a set of O II transitions and are generally blended, even for
stars that appear to rotate slowly. The selected spectral regions for this abundance
study are gathered in Table 2 where, in the first column, we indicate the
species that is dominant and whose abundance was adjusted in that 
particular region.  
Spectral lines used to construct linelists for the syntheses were taken from the 
Kurucz website (URL cfaku5.harvard.edu/) and the line  
gf-values were taken from the Opacity Project (OP), but also  
from Kurucz, when the atomic data needed was not available in the 
OP database. 

\subsection{LTE Abundances}

LTE abundances of carbon, nitrogen, oxygen, magnesium, aluminum, and silicon 
were determined by fitting synthetic spectra calculated with 
the program LINFOR (originally developed by H. Holweger, M. Steffen, \& W. Steenbock) 
to the observed spectral regions containing the selected individual transitions.
The synthetic profiles were then broadened for 
$v \sin i$ and limb darkening using a linear limb darkening coefficient
interpolated from the values from table 2 in Wade \& Rucinski (1985). In addition, 
the synthetic spectra were broadened by the instrumental profile, while the 
macroturbulent velocity was set to zero (as suggested by Ebbets 1979 
for main-sequence B-stars). As done in the abundance analyses of the 
sample of stars with low $v \sin i$, the O II transitions were used to derive 
microturbulent velocities as these lines are the most numerous in the sample 
and, most importantly, span a range
in line-strength adequate to constrain the microturbulence.
As a first step in this analysis, the O II line profiles were calculated 
for various values of microturbulence.  
The $v \sin i$\,s in each case
were allowed to vary and the best fit oxygen abundance
for each profile was determined by means of a $\chi ^2$-minimization.
Figure 1 illustrates the type of diagram constructed in order
to estimate the microturbulent velocity: this figure
displays the behavior of the O II lines for the sample star HD239745, 
where best fit oxygen abundances (corresponding to seven spectral regions
containing O~II lines) are calculated for microturbulent velocities 
ranging from 2 to 10 kms$^{-1}$. Such a diagram is the equivalent of
the usual `abundance versus equivalent width' plot, where a 
solution is found when 
zero slope is obtained. By definition, the solution in Figure 1 will be 
the microturbulence value that yields approximately the same
abundance for strong and weak lines: for this star a region of coincidence
is found around a microturbulent velocity of 8.0 kms$^{-1}$.

Once a value for the microturbulence was established for each star,
LTE synthetic spectra were calculated for all of the spectral regions listed
in Table 2. Here again the $v \sin i$\,s were left as free parameters
and adjusted in each spectral synthesis and an individual
projected rotational velocity was derived from each fit. 
The derived $v \sin i$\,s
for the individual fits did not, as expected, differ significantly and the
average values are presented in the third column of Table 3. 
The average LTE abundances for the studied elements,  
their standard deviations and 
the number [n] of fitted lines corresponding to each species are also found 
in Table 3. We stress that the carbon results in this table are for C~III,
while those in Papers I and II are from C II.  For one star in the 
sample (the Cep OB2 member HD239745), however, it was 
possible to compare the C II and C III abudances as, for this star,
the  C II lines at $\lambda$5143.50\AA, 5145.17\AA, and 5151.08\AA\
were strong enough
to be fitted: a carbon abundance of 8.22 provides a good fit 
for the 3 C II lines, as well as the C III line whose abundance
is listed in Table 3.
Such an agreement between C II and C III abundances is not always
achieved. In the literature, a few studies have pointed out that the
C III lines at 4650\AA\ usually provide abundances higher than the
C II lines (e.g. Smartt et al. 1996,  Rolleston et al. 1993). A discrepancy
is also found from inspection of the results given in Gies \& Lambert (1992).
In order to investigate systematic differences between C~II and C III
abundances, we decided to pursue an analysis of the C III lines 
in the stars analysed previously in Papers I and II and for which
we published C II abundances. The results of the comparison will be 
discussed in Section 4.1.   

The LTE abundances obtained follow trends found previously, 
namely that the elemental abundances in OB stars
are, in general, sub-solar. However, a more detailed discussion
must await results that take into account non-LTE effects.  Non-LTE 
calculations for C, N, O and Si have already been
presented in Papers I and II and proven to be small for the effective 
temperature and surface gravity range encompassed by the sample stars.
Non-LTE synthetic profiles were calculated for each observed profile, and,
in this study, new non-LTE calculations were added for two 
additional elements: Mg and Al.

\subsection{Non-LTE abundances}

Non-LTE synthetic profiles were calculated using 
model atoms  by Eber ( 1987 -- C~III),
Becker \& Butler (1989 -- N II), 
Becker \& Butler (1988 -- O II), Dufton {\it et al.} (1986 -- Al~III), 
Przybilla {\it et al.} (2001-- Mg II) and  Becker \& Butler 
(1990 -- Si III).
The adopted model atoms contain most of the energy levels belonging to
the main ionization stages with the addition of the lower levels of the
adjacent stages with relevant populations in this temperature range.
The completeness of the adopted model atoms is essential to guarantee an
accurate
ionization balance and the correct calculation of the level populations.
The models include a number of permitted bound-bound transitions
explicitly calculated, as well as transitions from levels fixed in LTE.

The level populations were computed 
with the program DETAIL, assuming LS-coupling to find the 
solution for the  equations of statistical equilibrium 
and transfer; the line profiles were computed with Voigt 
profile functions using the program SURFACE. As a first step
here, non-LTE synthetic profiles were  fitted to the observed O II lines 
in order to determine the microturbulent velocity in a similar manner as  
done for the LTE analysis and discussed in Section 3.1. 
The microturbulence derived from O II lines for each star was then
adopted to calculate the synthetic profiles for the species of the other
studied elements. 

The left panels in Figure 2 show synthetic spectra calculated for the region 
between $\lambda$4412 
and 4419\AA\ for HD239748. In this spectral region, the main contributions 
to the observed profiles
correspond to the O II lines at $\lambda$4414\AA\ and 4416\AA.
In the top left panel five synthetic profiles are shown 
for different oxygen abundances of log $\epsilon$(O) = 8.22, 8.37, 
8.42, 8.47 and 8.62 and for $v \sin i$ of 65 km$^{-1}$ (as indicated from
the $\chi^2$-minimization shown in the bottom right panel.)
The best fit oxygen abundance is represented by the synthetic profile 
with a solid line. It was derived via a 
$\chi^2$-minimization which
is shown in the top right panel of this same figure.
The bottom panel illustrates the sensitivity of the profiles to the 
variation of the projected rotational velocities, $v \sin i$, ranging, 
in this example, between
50 and 80 ${\rm km s}^{-1}$. 

Similarly to what was done in the LTE analysis,
a best fit non-LTE abundance and $v \sin i$ value were obtained for
each individually synthesized profile and Table 4 lists the average
abundances and $v \sin i$\,s for each target star. These are the 
final abundance results for C, N, O, Mg, Al and Si obtained in this study.
The derived $v \sin i$ values had a small variation from fit-to-fit 
with standard deviations 
typically smaller than 10kms$^{-1}$. Moreover, these $v \sin i$\,s
are very consistent with the $v \sin i$\,s derived from the LTE syntheses.
In order to investigate possible spurious trends due to systematic
errors in the adopted T$_{\rm eff}$ scale, the non-LTE
abundances obtained are plotted as a function of the effective
temperatures in Figure 3: no major trends are found.  
The derived microturbulences in non-LTE are in all cases smaller
than those derived in LTE, with the average being 2 km s$^{-1}$ less.
Since the microturbulence is an adhoc parameter needed in order to obtain
an agreement between the abundances of strong and weak lines, it might be 
expected that a more realistic non-LTE treatment would reduce 
the need for microturbulence. This seems to be the case, but just
slightly, as the derived microturbulences are still significant.
However, this abundance analysis is based upon one dimensional
plane-parallel model atmospheres in LTE and assume a constant microturbulent
velocity. These models are still far from being able to represent
the atmospheres of early-type stars perfectly and the microturbulence
remains a required parameter in these calculations. 

\subsection{Uncertainties}

The uncertainties in the adopted effective temperatures and surface 
gravities have been discussed in some detail in Paper I, with estimated 
errors of 4\% in T$_{\rm {eff}}$ 
and of 0.1 dex in $\log g$.  The uncertainty in the microturbulent 
velocity is assigned to be about $\pm1.5{\rm km s}^{-1}$.
The final abundances are the non-LTE results and it is the non-LTE
calculations that were used to investigate the final abundance uncertainties.
Non-LTE synthetic profiles were calculated for new model atmospheres
generated by adding the uncertainties to the adopted parameters for two
stars in the sample.
The elemental abundances derived
from profile fitting are also subject to uncertainties in the continuum
location as well as the choice of the projected 
rotational velocity.  
Based on the the variation of $\chi^2$ as a function of $v \sin i$, as
shown on the bottom panel of Figure 3, an error of 7\% is adopted due to 
$v \sin i$ and 5\% due to continuum location. 
We note that variations in the linear limb darkening coefficient have
negligible effects on both the derived values of $v \sin i$ and abundances.
The resulting abundance uncertainties, as well as the total expected error
($\delta_t$), are listed in Table 5.  
The total errors are $\sim$ 0.10--0.25 dex, except
for C III, which has a slightly larger error of $\sim$0.3 dex that 
is dominated by
the uncertainty in the effective temperature. The abundances of silicon 
and magnesium, based on strong lines, are very sensitive to the 
choice of the microturbulent velocity.

Previous abundances presented in Papers I and II were derived from
equivalent width measurements. In order to investigate any systematic
differences between abundances based on syntheses compared to those
based on equivalent widths, all stars analyzed in both Papers I and
II were re-analyzed using the synthesis techniques presented here.
No significant offsets in the abundances were found, with differences
typically less than 0.05 dex: both equivalent width and synthesis
based abundances can be compared directly.

\section{Discussion}

\subsection{The General Abundance Trends}

It is usual to compare stellar abundance results with the abundances
derived from the solar photosphere and meteorites.  The Sun
has been a reference point for obvious reasons. In this respect,
several abundance studies of OB stars in the Galactic disk have
noted the subsolar abundance values obtained for these early-type
stars.  With regard to this comparison, the non-LTE abundance results here 
indicate the same general pattern.  For nitrogen and oxygen the
derived target abundances are, in general, below the solar value. Some comments
should be made concerning the accepted solar oxygen abundance. Two
recent studies derive significantly lower solar O abundances than generally
accepted earlier values, with Prieto, Lambert \& Asplund (2001)
finding log $\epsilon$(O) = 8.69 and Holweger (2001) deriving 8.74. For the
time being, we simply average these recent determinations and use 
log $\epsilon$ (O) =8.72 as the reference solar value. This new O
abundance for the Sun also brings it closer to O abundances found in the
ISM, with log $\epsilon$ (O) = 8.43 being the gas phase abundance as
quoted by Sofia \& Meyer (2001): gas phase ISM abundances for many
elements are expected to be lower than stellar abundances for gas and
stars from similar populations as some material in the ISM is in the
form of grains. The lower solar abundances also result in better agreement
with B stars in general, although these stars still fall below even
this lower abundance. Subsolar abundances are also found for
Al, with the exception of one star for which a solar Al abundance
was derived. 
For Mg and Si, it is found that 
four stars are above, or at, the solar abundance level,
while the other stars lie below the solar abundance line. 
Figure 4 shows the non-LTE corrections for Mg II 
represented by $\Delta$(LTE - non-LTE). These are always negative and
can be as large as 0.25 dex. On the other hand, 
the LTE results would represent a very good approximation for aluminum,
where the differences between LTE and non-LTE Al III abundances scatter
around zero, with absolute values smaller than 0.08 dex (Figure 4). 
The corrections for nitrogen, oxygen and silicon are positive
for all studied stars and typically smaller than 0.2 dex.

For carbon the abundance results are close to the solar value but these
abundances need further discussion: 
for the 8 stars for which it was possible to
obtain C abundances (derived in all cases solely from C III),
4 of them have carbon abundances that are very close (within 0.05 dex)
to the solar value.
We note that the derived carbon abundances here are in general closer
to the solar abundance than other results in 
the literature, including our own in Papers~I and II.
This may be a systematic effect due to the choice of different lines
and, in particular, different ionization stages: here strong C III lines
and in Papers I and II the weaker C II transitions. Non-LTE
C III corrections with respect to LTE abundances are not very large and
range between $-0.16$ to +0.06 dex. These corrections, shown in Figure 4,
are not able to bring the C II and C III into agreement. The non-LTE
corrections for C II are typically smaller than 0.1 dex. A
comparison of the C III and C II abundances was possible for 8 stars from
Papers I and II. The derived 
results indicate that the LTE C III abundances are higher.
For two stars we see very large discrepancies
between the C III and C II results (both in LTE and non-LTE) of the
order of 0.7--0.9 dex.  Clearly further work is needed to clarify the
situation and this is in progress.

Taken together,
the mean non-LTE abundances for this sample of B stars still point
to a general pattern of sub-solar abundances.  Average abundances are
[C/H]= $-0.15\pm$0.15, [N/H]= $-0.44\pm$0.15, [O/H]= $-0.22\pm$0.14,
[Mg/H]= $-0.13\pm$0.24, [Al/H]= $-0.41\pm$0.21, and [Si/H]= $-0.25\pm$0.30.
The overall average abundance for all 6 elements, relative to solar, is
[m/H]= $-0.27\pm$0.13, or about a factor of 2 lower in metallicity than
the Sun. 
Although subsolar, these abundances are still larger than most of the gas phase
ISM abundances as summarized recently by Sofia \& Meyer (2001). 
Two elements, N and Si, are found to be below the gas phase ISM abundances.
We note, however, that even for N in the Sun this seems to be marginally the
case as well, while Si has large abundance uncertainties, 
as noted earlier.  

\subsection{ Cep OB2} 

Nine stars of this sample belong to the Cep OB2 association, 
which has already been analyzed chemically in Paper I by means of 
equivalent-width measurements of spectral lines in sharp lined stars.
The mean non-LTE abundances for a sample of 8 low $v \sin i$ stars 
are $\log \epsilon$ (C)=8.17$\pm$0.09, $\log \epsilon$ (N)=7.62$\pm$0.10, 
$\log \epsilon$ (O)=8.61$\pm$0.10 and $\log \epsilon$ (Si)=7.21$\pm$0.28.
This study  may now be complemented with the present sample, for which it 
is found to have mean non-LTE abundances of 8.37$\pm$0.15 for carbon,
7.53$\pm$0.17 for nitrogen, 
8.47$\pm$0.14 for oxygen and 7.28$\pm$0.33 for silicon. The average
abundances derived from both samples agree quite well for N, O and Si,
despite the somewhat higher dispersions obtained for the sample of high 
$v \sin i$ stars.  Carbon abundances are somewhat larger in this sample
than in Paper I. However, as discussed before, carbon results 
in this paper are based 
on C III, instead of C II (as in Paper I): part of this difference may
thus reside in systematic effects between C II and C III. 
If both samples are combined, then the entire set consists of 
17 stars belonging to the Cep OB2 association and this represents $\sim$ 30 
percent
of all the stars formed in the Cep OB2 association with spectral types 
between O9 and B3. These average abundances are then, relative to solar,
[C/H]= $-0.26$, [N/H]= $-0.40$, [O/H]= $-0.18$, and [Si/H]=$-0.30$: all underabundant
by roughly the same amount.
The mean abundances confirm the results of Cep OB2
being slightly metal poor relative to the Sun by about 0.2-0.3 dex (the average
underabundance of the four elements discussed here is $-0.28$).

With the addition of the sample of high $v \sin i$ stars, the number of   
Cep OB2 stars is twice as large as the number presented in Paper I and
provides tighter constraints on the chemical distribution in the association. 
Cep OB2 is believed to be divided into two subgroups of different ages:
Cep OB2a, with an estimated age of $7 \times 10^6$ years, extending over
a large area within the association, and Cep OB2b, with an estimated age of 
$3 \times 10^6$ years, associated with the open cluster Trumpler 37 (Tr 37).
It might be assumed that those stars assigned as members of Tr 37 by 
Garmany \& Stencel (1992) belong to the younger subgroup (HD\,s 205794, 
206183, 206267D, 207538, 239724, 205948, 239729, and 239748),
while the rest of the sample (HD\,s 206327, 239742, 239743, 207951, 209339,  
235618, 239681, 239710 and 239745) belongs to the older subgroup, Cep OB2a.

This assumption can be tested by using the derived stellar parameters to
construct a variation of an HR-diagram for Cep OB2.  The surface gravity
(as log g) is plotted versus log T$_{\rm eff}$ and this is shown in the
top panel of Figure 5 for all 17 Cep OB2 members in the sample.  The open
symbols correspond to members assigned to Tr 37 (and would be expected to
be the younger stars of Cep OB2b), while the filled symbols are the other
Cep OB2 stars.  Stellar model tracks from Schaller et al. (1992) are
shown as solid curves for M= 7M$_{\odot}$, 9M$_{\odot}$, 12M$_{\odot}$,
15M$_{\odot}$, and 20M$_{\odot}$ models (with solar
heavy-element abundances), while the dashed line shows the zero-age main
sequence (ZAMS).  It is clear from this form of an HR-diagram that the
Cep OB2 members in this sample do generally segregate into two
evolutionary groups, with a set of 10 stars falling very near the 
ZAMS, and a second, somewhat evolved group of 7 stars.  The earlier
assumption of assigning age status based on Tr 37 membership is a reasonable
one: 7 of 8 stars identified as Tr 37 members fall on the ZAMS.  Using the
HR-diagram as the discriminator, 7 stars should be put into the older
group (Cep OB2a: HD\,s 206327, 207951, 235618, 239681, 239724, 239742,
and 239743), and 10 stars into the younger group (Cep OB2b: HD\,s 205794,
205948, 206183, 206267D, 207538, 209339, 239710, 239729, 239745, and 239748). 

Abundances can be compared between the two subgroups of Cep OB2 (again
using only the non-LTE abundances).  Average abundances for Cep OB2a
(the older subgroup) are $\log \epsilon$ (C)= 8.27$\pm$0.20, 
$\log \epsilon$ (N)= 7.66$\pm$0.08, $\log \epsilon$ (O)= 8.60$\pm$0.13,
and $\log \epsilon$ (Si)= 7.28$\pm$0.32 (Mg and Al are not included in the
subgroup comparison as non-LTE results are not available from Paper I).
The corresponding average abundances for Cep OB2b are 
$\log \epsilon$ (C)= 8.32$\pm$0.15,
$\log \epsilon$ (N)= 7.52$\pm$0.15, $\log \epsilon$ (O)= 8.49$\pm$0.14,
and $\log \epsilon$ (Si)= 7.23$\pm$0.31.  The two subgroups have
similar average abundances considering the abundance uncertainties, and
this confirms the conclusions from Paper I that Cep OB2 is rather
chemically homogeneous.  

The behavior of the N/O abundance ratios are investigated more
carefully in the bottom two panels of Figure 5, where the
log(N/O) abundances are plotted versus log g (middle panel) and
$v \sin i$ (bottom panel).  The abundance ratio of N/O is shown, as
the ratio is somewhat less sensitive to stellar parameter uncertainties,
while surface gravity is used as a measure of evolutionary state (see
the top panel of Figure 5), and projected rotational velocity is used to
test for rotationally-induced mixing (Heger \& Langer 2000).  In general,
the N/O ratios of stars in both subgroups display small scatter, with
log(N/O)= $-0.96\pm$0.13, however, two N-rich stars do stand out, especially
in the middle panel as having lower values of log g (and are thus more
evolved).  These two stars are HD235618 and HD239681 and are Cep OB2a
members.  In the top panel of Figure 5, these N-rich stars are the two
most evolved stars falling nearly on the 15M$_{\odot}$ model track and
also have two of the largest projected rotational velocities (bottom
panel of Figure 5). It should be remembered, of course, that stars with
low $v \sin i$ could also be rapid rotators viewed nearly pole-on, but
stars with high $v \sin i$ values are rapid rotators, thus HD235618 and
HD239681 are rotating rapidly. This suggests that rotation in fairly massive stars
has led to mixing of CN-processed material to the surface.  Nitrogen
enhancements of nearly a factor of two are indicated.  In a simple mixing
scheme of only CN-processing, the sum of C plus N nuclei will be
conserved; with an assumed starting abundance of $\log \epsilon$ (C)= 8.35
and $\log \epsilon$ (N)= 7.55, then an increase to $\log \epsilon$ (N)= 7.80
(as observed), would result in a carbon abundance change of only $-0.08$ dex:
such a change in C would be lost in the noise of the C-abundance
uncertainties (especially in view of the possible differences between
C II and C III abundances).  

Heger \& Langer (2000) have computed models of rotating massive stars
(M= 8--20M$_{\odot}$) that include surface abundance changes caused by
rotationally induced mixing.  They find that surface $^{14}$N-abundances
can increase by about a factor of 2 (as observed in two of the Cep OB2
members), in 12--20M$_{\odot}$ stars that have rotational velocities
greater than about 200 km s$^{-1}$ and ages of $\sim$2--5 Myr.  The
two Cep OB2 stars with N-abundances enhanced by about +0.3 dex have masses 
close to 
15M$_{\odot}$, and ages of a few Myr.  With projected rotational velocities
of 100 and 140 km s$^{-1}$, respectively, these two stars could easily have
true rotational velocities of 200 km s$^{-1}$ or larger.
A test of the mixing hypothesis could be
facilitated by a comparison of boron abundances in a sample of these stars. 

\section{Conclusions}

Both the LTE and non-LTE abundances of C, N, O, Mg, Al, and Si derived
here continue to indicate that young O and B stars in this part of the
Galaxy have slightly subsolar abundances.  
The inclusion of C III lines
in this analysis does, however, yield carbon abundances that are
somewhat closer to solar than those abundances derived from C II lines.

With 9 Cep OB2 members in this sample, plus 8 members with low values of $v \sin i$
studied previously (Paper I), the chemical abundances in this
association are now better defined and we have 
probed the chemical composition of the gas that formed the OB stars
of this association with stellar masses roughly between 7--15 M$_{\odot}$ as much possible.
Two distinct age subgroups within
Cep OB2 can be identified based upon the stellar parameters derived here,
with both subgroups having similar elemental abundances.  Two N-rich
stars are identified in the older subgroup, however, and these two
stars are both rapidly rotating ($v \sin i \ge$ 100 km s$^{-1}$) and
rather massive (M$\sim$ 15 M$_{\odot}$).  Their nitrogen overabundances
compare well to those predicted from stellar models by Heger \& Langer (2000)
that include surface abundance changes brought about by rotationally 
induced mixing.

S.D. acknowledges a DAAD fellowship and the people at the 
Sternwarte, Munich.  K.C. thanks David Lambert for travel support to observe
at McDonald Observatory.
V.V.S. acknowledges support from NASA (NAG5-9213) and the National
Science Foundation (AST99-87374).


\clearpage
\begin{table}
\centering
\small
\textsc{Table 1. Atmospheric Parameters}

\vspace{0.5cm}
\footnotesize
\label{}
\begin{tabular}{c c l l}
\hline
Star & Association  & T$_{\rm eff}$(K)  & $\log g$ \\
\hline
HD 202347 & Cyg OB7 &  23280 ${^b}$ & 4.13 ${^b}$ \\
HD 205948 & Cep OB2 &  24350 ${^a}$ & 4.25 ${^a}$ \\ 
HD 207951 & Cep OB2 &  20650 ${^a}$ & 3.88 ${^a}$ \\
HD 209339 & Cep OB2 &  31250 ${^a}$ & 4.28 ${^a}$ \\
HD 227696 & Cyg OB3 &  29100 ${^b}$ & 4.45 ${^b}$ \\
HD 228199 & Cyg OB3 &  29870 ${^b}$ & 4.45 ${^b}$ \\
HD 235618 & Cep OB2 &  27180        & 3.75        \\ 
HD 239681 & Cep OB2 &  26830        & 3.70        \\
HD 239710 & Cep OB2 &  21900        & 4.50        \\ 
HD 239729 & Cep OB2 &  28450 ${^a}$ & 4.22 ${^a}$ \\ 
HD 239745 & Cep OB2 &  27340        & 4.45        \\ 
HD 239748 & Cep OB2 &  27480        & 4.42        \\
\hline
\end{tabular}
\footnote{$^a$  Paper I; $^b$ Paper II. \\}
\end{table}

\clearpage
\begin{table}
\centering
\small
\textsc{Table 2. Linelists}

\vspace{0.5cm}
\footnotesize
\label{}
\begin{tabular}{c l c c c }
\hline
Wavelength interval & $\lambda$(\AA) & Species & $\chi$(eV) & $\log (gf)$ \\
\hline
4234-4244 & 4236.93 & N II  & 23.24 & $\,\,$0.39 \nl
 N II     & 4237.05 & N II  & 23.24 & $\,\,$0.56 \nl
          & 4237.94 & O II  & 28.83 & -0.99 \nl
          & 4239.48 & O III & 33.15 & -2.04 \nl
          & 4241.75 & N II  & 23.24 & $\,\,$0.22 \nl
          & 4241.79 & N II  & 23.25 & $\,\,$0.72 \nl
          & 4242.50 & N II  & 23.25 & -0.34 \nl
\hline
4412-4419 & 4413.11 & O II  & 28.94 & -0.73 \nl
  O II    & 4414.88 & O II  & 23.44 & $\,\,$0.22 \nl
          & 4416.97 & O II  & 23.42 & -0.04 \nl
          & 4417.10 & N II  & 23.42 & -0.34 \nl
          & 4418.84 & S III & 18.24 & -1.92 \nl
\hline
4448-4455 & 4448.34 & O II  & 28.36 & $\,\,$0.07 \nl
O II      & 4452.38 & O II  & 23.44 & -0.73 \nl
\hline
4476-4484 & 4479.88 & Al III & 20.78 & 0.90 \nl
Mg II, Al III & 4479.97  & Al III & 20.78 & 1.02 \nl
          & 4481.13 & Mg II & 8.86 & 0.74 \nl
          & 4481.15 & Mg II & 8.86 & -0.56 \nl
          & 4481.33 & Mg II & 8.86 & 0.59 \nl
\hline
4550-4554 & 4552.41 &  S II & 15.07 & -0.10 \nl
  Si III  & 4552.62 & Si III& 19.02 & $\,\,$0.28 \nl
\hline
4565-4570 & 4567.84 & Si III& 19.02 &  $\,\,$0.06 \nl
 Si III   & 4569.06 & Ne II & 34.93 &  $\,\,$0.14 \nl
          & 4569.26 & O III & 46.00 &  $\,\,$0.07 \nl
\hline
4572-4576 & 4574.42 & Ne II & 34.84 & -0.16 \nl
 Si III   & 4574.76 & Si III& 19.02 & -0.42 \nl
\hline
4589-4593 & 4591.01 & O II  & 25.66 &  $\,\,$0.32 \nl
O II      &         &       &       &       \nl
\hline
4605-4614 & 4607.15 & N II  & 18.46 & -0.48 \nl
N II, O II& 4608.08 & N II  & 23.48 & -0.25 \nl
          & 4609.37 & O II  & 29.07 &  $\,\,$0.71 \nl
          & 4610.17 & O II  & 29.06 & -0.17 \nl
          & 4610.61 & O III & 45.94 & 0.02 \nl
          & 4613.10 & O II  & 29.07 & -0.59 \nl
\hline
4628-4632 & 4629.97 & C II  & 24.79 &  $\,\,$0.50 \nl
N II      & 4630.54 & N II  & 18.48 &  $\,\,$0.09 \nl
          & 4631.27 & Si IV & 36.42 &  $\,\,$0.85 \nl
          & 4631.27 & Si IV & 36.42 & -0.58 \nl
          & 4631.27 & Si IV & 36.42 &  $\,\,$0.96 \nl
\hline
4637-4545 & 4638.28 & Si III& 28.07 & -0.44 \nl
N II, O II& 4638.86 & O II  & 22.97 & -0.35 \nl
          & 4640.64 & N III & 30.46 &  $\,\,$0.14 \nl
          & 4641.83 & O II  & 22.98 &  $\,\,$0.05 \nl
          & 4641.85 & N III & 30.46 & -0.81 \nl
          & 4643.09 & N II  & 18.48 & -0.39 \nl
\hline
4647-4651 & 4647.42 & C III & 29.54 &  $\,\,$0.06 \nl
C III, O II & 4647.59 & O II  & 29.06 & -0.64 \nl
          & 4649.14 & O II  & 23.00 &  $\,\,$0.33 \nl
          & 4650.25 & C III & 29.54 & -0.15 \nl
          & 4650.85 & O II  & 23.00 & -0.35 \nl
          & 4651.02 & C III & 38.22 & -0.47 \nl
          & 4651.47 & C III & 29.54 & -0.63 \nl
\hline
\end{tabular}
\end{table}

\clearpage
\begin{table}
\centering
\small
\textsc{Table 2. Linelists ({\it continued})}

\vspace{0.5cm}
\footnotesize
\label{}
\begin{tabular}{c l c c c }
\hline
Wavelength interval & $\lambda$(\AA) & Species & $\chi$(eV) & $\log (gf)$ \\
\hline
4658-4664 & 4659.06 & C III & 38.22 & -0.69 \nl
O II      & 4661.64 & O II  & 22.98 & -0.25 \nl
          & 4663.64 & C III & 38.22 & -0.57 \nl
          & 4665.86 & C III & 38.22 & 0.01 \nl
\hline
4904-4908 & 4904.78 & N III & 39.40 & -0.26 \nl
 O II     & 4906.82 & O II  & 26.30 & -0.05 \nl
\hline
4939-4945 & 4941.10 & O II  & 26.55 &  $\,\,$0.07 \nl
 O II     & 4943.00 & O II  & 26.56 &  $\,\,$0.33 \nl                      
\hline
5003-5012 & 5005.15 & N II  & 20.67 &  $\,\,$0.61 \nl
 N II     & 5007.33 & N II  & 20.94 &  $\,\,$0.17 \nl
          & 5010.62 & N II  & 18.47 & -0.61 \nl
\hline
\end{tabular}
\end{table}

\clearpage
\begin{table}
\centering
\small
\textsc{Table 3. LTE Abundances }

\vspace{0.5cm}
\hspace{0.3cm}
\footnotesize
\label{}
\begin{tabular}{ccccccccc}
\hline
Star & $\xi$ & $v\sin i$ & $\log \epsilon({\rm C})$ & $\log \epsilon({\rm N})$ &
$\log \epsilon({\rm O})$ & $\log \epsilon({\rm Mg})$ & $\log \epsilon({\rm Al})$ &
$\log \epsilon({\rm Si})$  \\
   & ${\rm (km s^{-1})}$  & ${\rm (km s^{-1})}$  &  &  & & & & \\
\hline
HD 202347 &7.5& 121$\pm$7 & -  & 7.67$\pm$0.10 [3]& 8.62$\pm$0.11 [5]& 7.15 [1]& 6.30 [1]& 7.28$\pm$0.04 [3]\\
HD 205948 & 7 & 146$\pm$12& 8.27 [1]& 7.53$\pm$0.09 [2]& 8.47$\pm$0.10 [4]& 6.98 [1]& 6.20 [1]& 7.13$\pm$0.13 [3]\\
HD 207951 &6.5&  84$\pm$2 & 8.53 [1]& 7.76$\pm$0.03 [3]& 8.75$\pm$0.09 [5]& -  & -   & 7.18$\pm$0.06 [3]\\
HD 209339 & 7 & 101$\pm$9 & -  & 7.47$\pm$0.13 [3]& 8.55$\pm$0.14 [5]& 7.32 [1]& 6.10 [1]& 7.63$\pm$0.13 [3]\\
HD 227696 & 12& 120$\pm$5 & -  & 7.58$\pm$0.09 [2]& 8.62$\pm$0.11 [5]& - & -  & 7.61$\pm$0.14 [3]\\
HD 228199 & 8 & 105$\pm$5 & 8.44 [1]& 7.64$\pm$0.09 [3]& 8.77$\pm$0.14 [5]& 7.54 [1]& 6.10 [1]& 7.62$\pm$0.01 [3]\\
HD 235618 & 12& 101$\pm$2 & 8.60 [1]& 7.90    [1]& 8.67$\pm$0.11 [4]& 7.33 [1]& 6.00 [1]& 7.95$\pm$0.14 [2]\\
HD 239681 & 10& 140$\pm$10& 8.08 [1]& 7.73$\pm$0.15 [3]& 8.61$\pm$0.09 [5]& 7.70 [1]& 6.40 [1]& 7.83$\pm$0.18 [2]\\
HD 239710 & 8 &  63$\pm$6 & 8.40 [1]& 7.62$\pm$0.07 [3]& 8.63$\pm$0.09 [5]& 7.44 [1]& 5.95 [1]& 7.88$\pm$0.20 [3]\\
HD 239729 & 6 & 100$\pm$7 & -   & 7.44$\pm$0.11 [3]& 8.49$\pm$0.20 [7]& 7.19 [1]& 5.90 [1]& 7.07$\pm$0.02 [3]\\
HD 239745 & 8 &  61$\pm$3 & 8.22 [1]& 7.58$\pm$0.07 [3]& 8.57$\pm$0.05 [7]& 7.20 [1]& 5.80 [1]& 7.16$\pm$0.12 [3]\\
HD 239748 & 6 &  66$\pm$3 & 8.25 [1]& 7.68$\pm$0.09 [4]& 8.63$\pm$0.09 [7]& 7.35 [1]& 5.97 [1]& 7.31$\pm$0.06 [3]\\ 
\hline
\end{tabular}
\end{table}

\clearpage
\begin{table}
\centering
\small
\textsc{Table 4. Non-LTE Abundances}

\vspace{0.5cm}
\footnotesize
\label{}
\begin{tabular}{ ccccccccc  }
\hline
Star & $\xi$ & $v\sin i$ & $\log \epsilon({\rm C})$ & $\log \epsilon({\rm N})$ &
$\log \epsilon({\rm O})$ & $\log \epsilon({\rm Mg})$ & $\log \epsilon({\rm Al})$ &
$\log \epsilon({\rm Si})$  \\
   & ${\rm (km s^{-1})}$  & ${\rm (km s^{-1})}$  &  &  & & & & \\
\hline
HD 202347 & 6 & 119$\pm$6 & -    & 7.54$\pm$0.04 & 8.54$\pm$0.12 & 7.35 & 6.28 & 7.13$\pm$0.06 \\
HD 205948 & 5 & 144$\pm$11& 8.35 & 7.35$\pm$0.05 & 8.29$\pm$0.03 & 7.05 & 6.18 & 6.95$\pm$0.13 \\
HD 207951 & 5 &  87$\pm$2 & 8.47 & 7.65$\pm$0.05 & 8.72$\pm$0.10 &  -   &  -   & 7.05$\pm$0.07 \\
HD 209339 &3.5&  98$\pm$9 & -    & 7.34$\pm$0.03 & 8.36$\pm$0.12 & 7.38 & 6.07 & 7.44$\pm$0.03 \\    
HD 227696 & 10& 120$\pm$5 & -    & 7.49$\pm$0.06 & 8.60$\pm$0.16 &  -   &  -   & 7.41$\pm$0.16 \\
HD 228199 & 5 & 104$\pm$6 & 8.55 & 7.56$\pm$0.16 & 8.67$\pm$0.16 & 7.78 & 6.13 & 7.57$\pm$0.03 \\
HD 235618 & 9 & 100$\pm$3 & 8.54 & 7.81          & 8.47$\pm$0.15 & 7.55 & 5.94 & 7.73$\pm$0.02 \\
HD 239681 & 9 & 142$\pm$11& 8.10 & 7.70$\pm$0.16 & 8.48$\pm$0.17 & 7.82 & 6.48 & 7.55$\pm$0.03 \\
HD 239710 & 8 &  64$\pm$5 & 8.50 & 7.57$\pm$0.08 & 8.60$\pm$0.13 & 7.63 & 5.96 & 7.71$\pm$0.05 \\
HD 239729 & 5 &  99$\pm$8 & -    & 7.33$\pm$0.05 & 8.28$\pm$0.10 & 7.26 & 5.86 & 6.86$\pm$0.07 \\
HD 239745 & 6 &  61$\pm$2 & 8.38 & 7.51$\pm$0.13 & 8.46$\pm$0.09 & 7.25 & 5.77 & 7.08$\pm$0.19 \\
HD 239748 & 5 &  68$\pm$4 & 8.27 & 7.55$\pm$0.07 & 8.54$\pm$0.08 & 7.42 & 5.95 & 7.13$\pm$0.13 \\
\hline
\end{tabular}
\end{table}

\clearpage
\begin{table}
\centering
\small
\textsc{Table 5. Abundance Uncertainties.}

\vspace{0.5cm}
\footnotesize
\label{}
\begin{tabular}{ cccc }
\hline
Ion  &  correction & HD 239710 &  HD 239745 \\
\hline
C III & $\delta(T_{eff})$   & \,\,\,-0.25 & \,\,\,-0.28 \\
       & $\delta (\log g)$   & +0.07 & +0.08 \\
       & $\delta(\xi)$       & \,\,\, 0.00 & \,\,\,-0.06 \\
       & $\delta(v \sin i)$   & +0.03 & +0.03 \\
       & $\delta(\rm continuum)$ & +0.02 & +0.04 \\
       & $\delta_t$          & +0.26 & +0.30 \\   
\noalign{\hrule}
  N II & $\delta(T_{eff})$   & \,\,\,-0.10 & +0.05 \\
       & $\delta (\log g)$   & +0.03 & \,\,\,-0.01 \\
       & $\delta(\xi)$       & \,\,\,-0.05 & \,\,\,-0.01 \\      
       & $\delta(v \sin i)$   & +0.02 & +0.01 \\
       & $\delta(\rm continuum)$ & +0.02 & +0.03 \\
       & $\delta_t$          & +0.12 & +0.06 \\    
\noalign{\hrule}
  O II & $\delta(T_{eff})$   & \,\,\,-0.15 & \,\,\,-0.05 \\
       & $\delta (\log g)$   & +0.03 & +0.05 \\
       & $\delta(\xi)$       & \,\,\,-0.07 & \,\,\,-0.10 \\      
       & $\delta(v \sin i)$   & +0.02 & +0.01 \\
       & $\delta(\rm continuum)$ & +0.03 & +0.03 \\
       & $\delta_t$          & +0.17 & +0.13 \\  
\noalign{\hrule}
Mg II  & $\delta(T_{eff})$   & +0.13 & +0.12 \\
       & $\delta (\log g)$   & \,\,\,-0.07 & \,\,\,-0.07 \\
       & $\delta(\xi)$       & \,\,\,-0.16 & \,\,\,-0.07 \\    
       & $\delta(v \sin i)$   & +0.05 & +0.04 \\
       & $\delta(\rm continuum)$ & +0.03 & +0.05 \\
       & $\delta_t$          & +0.22 & +0.17 \\
\noalign{\hrule}
Al III & $\delta(T_{eff})$   & \,\,\,-0.01 & +0.03 \\
       & $\delta (\log g)$   & +0.02 & +0.03 \\
       & $\delta(\xi)$       & \,\,\,-0.04 & \,\,\,-0.04 \\   
       & $\delta(v \sin i)$   & +0.01 & +0.01 \\
       & $\delta(\rm continuum)$ & +0.01 & +0.02 \\
       & $\delta_t$          & +0.05 & +0.06 \\       
\noalign{\hrule}
Si III & $\delta(T_{eff})$   & \,\,\,-0.17 & +0.04 \\
       & $\delta (\log g)$   & +0.04 & \,\,\,-0.02 \\
       & $\delta(\xi)$       & \,\,\,-0.17 & \,\,\,-0.14 \\      
       & $\delta(v \sin i)$   & +0.02 & +0.02 \\
       & $\delta(\rm continuum)$ & +0.02 & +0.03 \\
       & $\delta_t$          & +0.24 & +0.15 \\    
\hline
\end{tabular}
\end{table}

\clearpage
\centerline{Figure Captions}

{Fig 1. Variation of LTE oxygen abundance with
the microturbulence velocity for the sample star HD 239745.
The adopted solution is $\xi=8{\rm km s^{-1}}$ that corresponds to
$\log \epsilon(O)=8.57\pm0.05$.}

{Fig 2. Comparison between observed and non-LTE synthetic
profiles for one of the spectral regions containing O II lines. The observed
spectrum is for the target star HD239748 and the synthetic profiles were
calculated for different sets of parameters.
{\it Top}: on the left panel we show synthetic profiles calculated for
five values of oxygen abundances (indicated in the figure).
The best fit oxygen abundance is derived for $\log \epsilon (O)$=8.42
(represented by the solid line).
In the right panel we present the variation of $\chi^2$ as
a function of oxygen abundance.
{\it Bottom}: same for different values of  $v \sin i$
varying from 50 to 80 ${\rm km s}^{-1}$.}

{Fig 3. Non-LTE abundances as a function of 
adopted effective temperatures. The derived abundances show almost no
trend with effective temperature and are, in general, sub-solar. There
are some stars that have solar, and in some cases, above solar abundances 
of C, O, Mg, Al, and
Si. The reference solar abundances (dotted lines) are from Grevesse, Noels \& 
Sauval (1996) for all elements except for oxygen for which we adopt an
average abundance value between the recent studies by Prieto et al. (2001)
and Holweger (2001).} 

{Fig 4. The non-LTE correction $\Delta$(LTE - non-LTE)
as a function of effective temperature.}

{Fig 5. The derived stellar parameters T$_{\rm eff}$ and
log g are plotted in the top panel to form a modified HR-diagram for Cep OB2
stars.  Members of the young cluster Tr 37 are plotted as the open
circles, and all other association members are plotted as filled circles. 
The Tr 37 members should be generally younger and this is found to be the
case.  Solid curves are stellar models from Schaller et al. (1992) for
7, 9, 12, 15 and 20M$_{\odot}$ models, with the dashed line denoting
the ZAMS.  Two fairly well-defined, differing age subgroups are
identified (based upon the Schaller et al. models, the age difference is
$\sim$2--5 Myr).  The middle panel shows log(N/O) abundance ratios versus
surface gravity, as the surface gravity of a star will initially decrease
as it evolves off of the main sequence.  Note that the two highest N/O-ratio 
stars have the lowest surface gravities (these two are the most highly
evolved of the $\sim$15M$_{\odot}$ stars).  The dotted line is the average
log(N/O)=$-0.99$ of the other Cep OB2 stars.  The bottom panel plots
log(N/O) versus $v \sin i$ and shows that the two N-rich evolved stars
are also two of the most rapidly rotating ones: rotationally induced
mixing of CN-cycle material to the surface is one viable explanation for the
increased N/O ratios.}

\clearpage
\begin{figure}
\figurenum{1}
\centerline{\psfig{figure=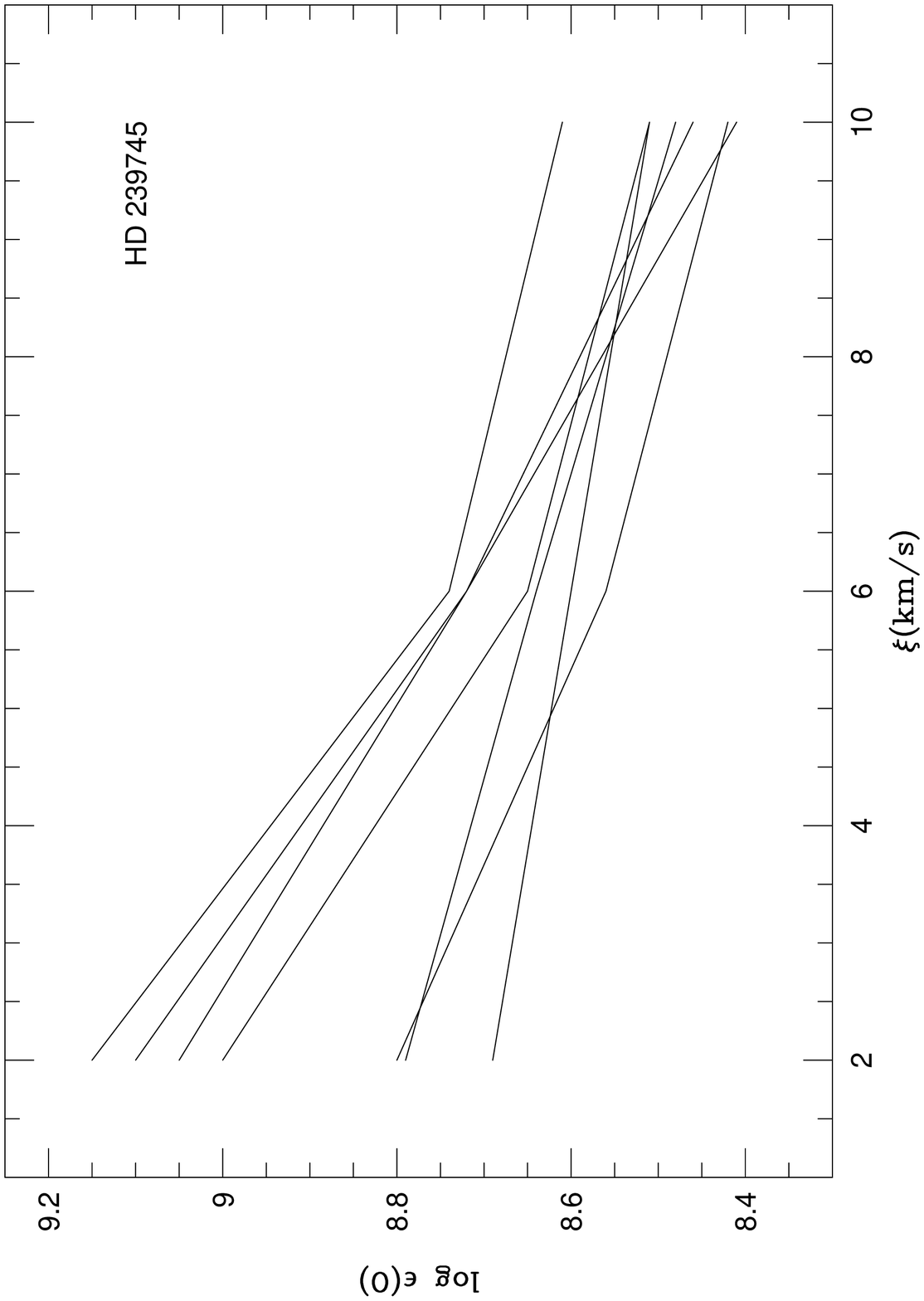,width=16cm,height=22cm} }
\caption{ }
\end{figure}

\clearpage
\begin{figure}
\figurenum{2}
\centerline{\psfig{figure=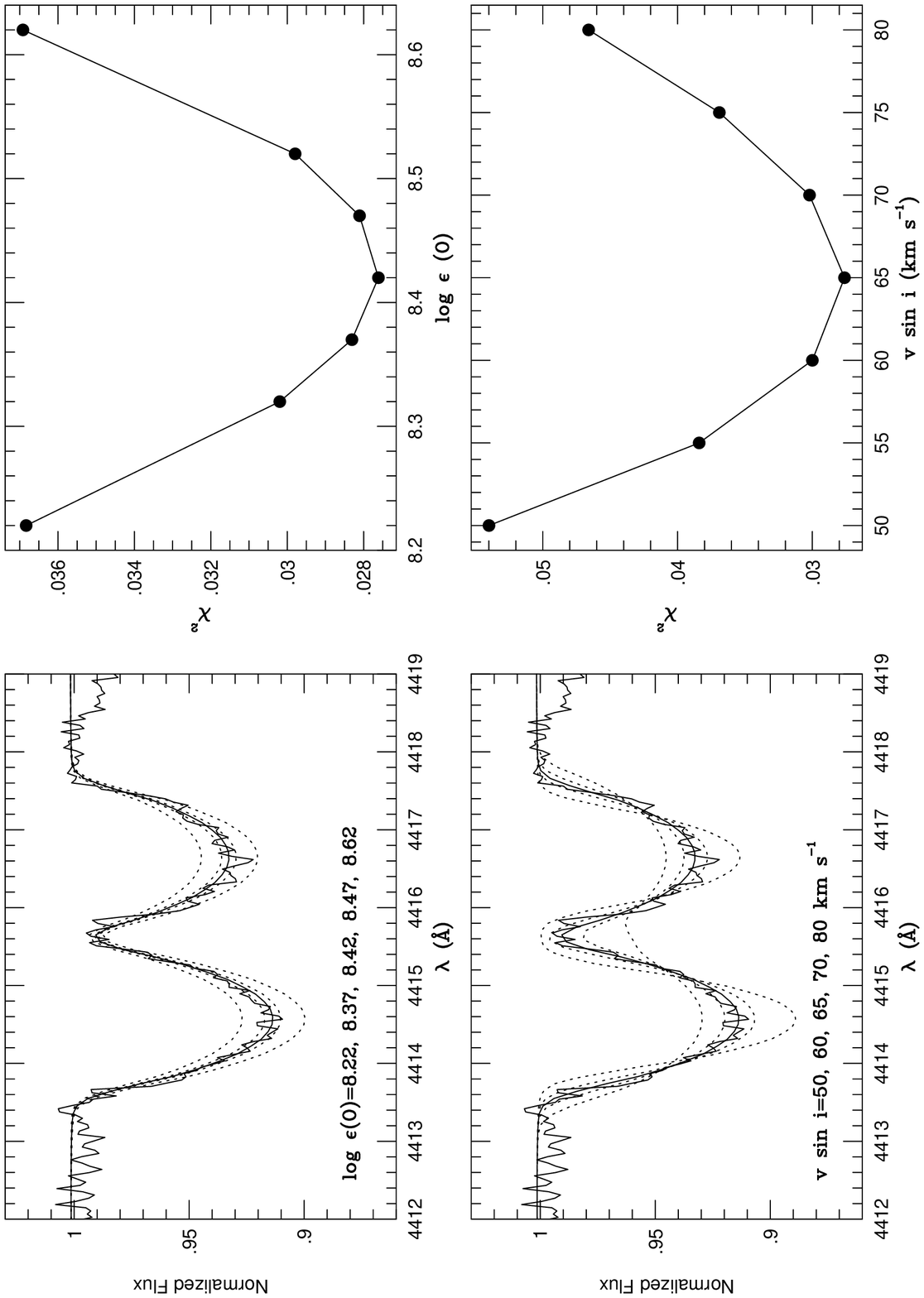,width=16cm,height=22cm} }
\caption{ }
\end{figure}

\clearpage
\begin{figure}
\figurenum{3}
\centerline{\psfig{figure=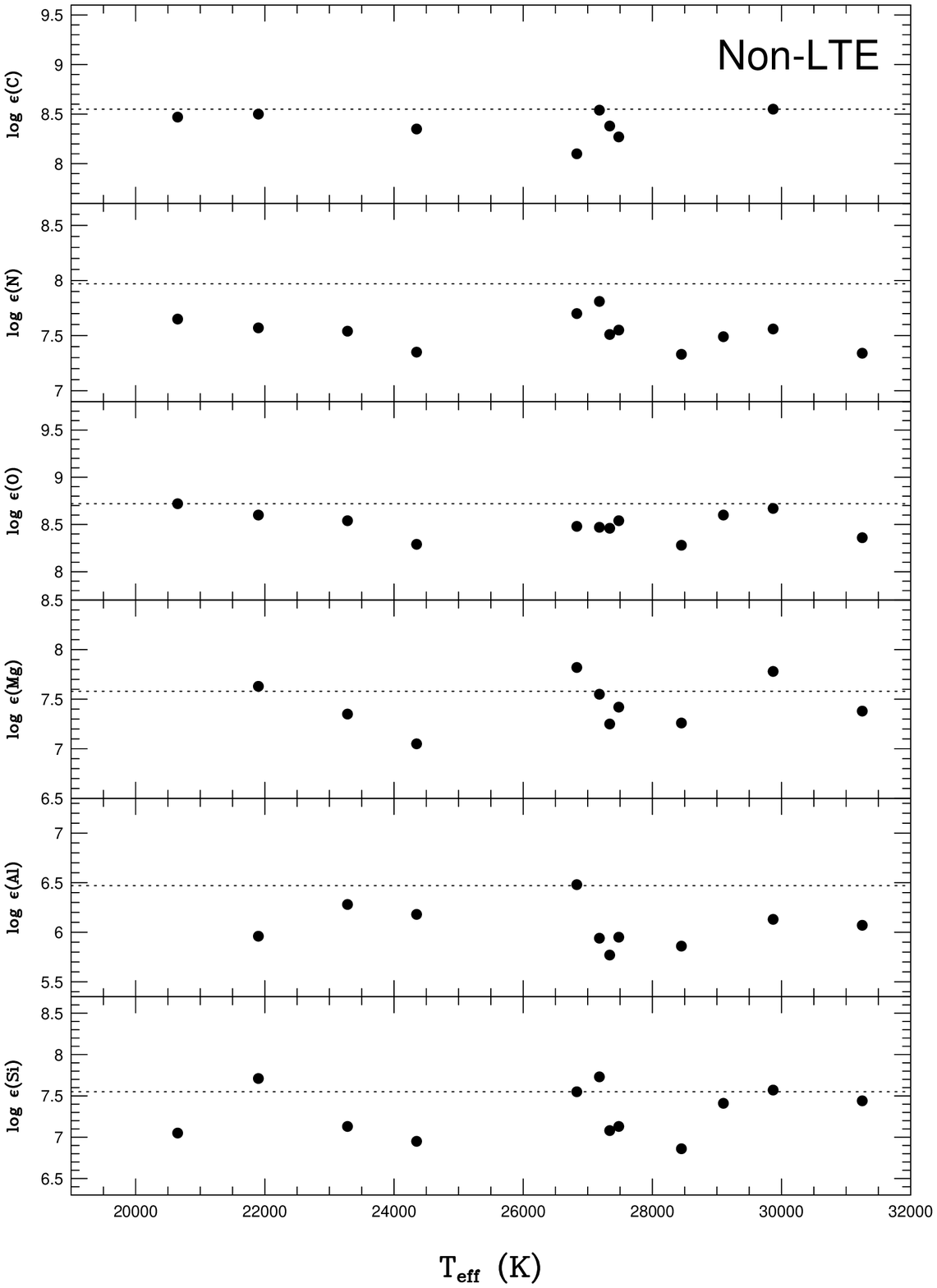,width=16cm,height=22cm} }
\caption{ }
\end{figure}

\clearpage
\begin{figure}
\figurenum{4}
\centerline{\psfig{figure=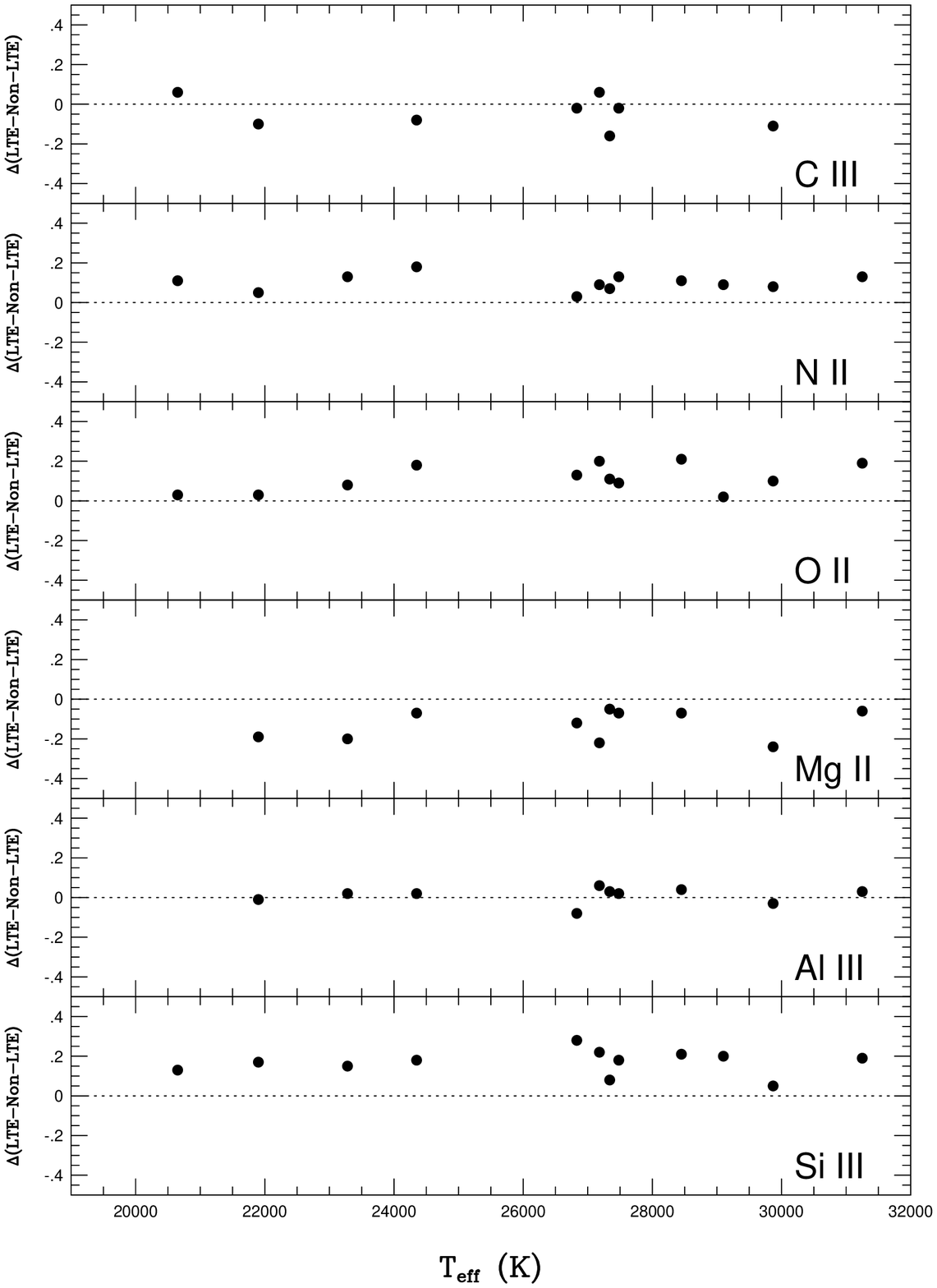,width=16cm,height=22cm} }
\caption{ }
\end{figure}

\clearpage
\begin{figure}
\figurenum{5}
\centerline{\psfig{figure=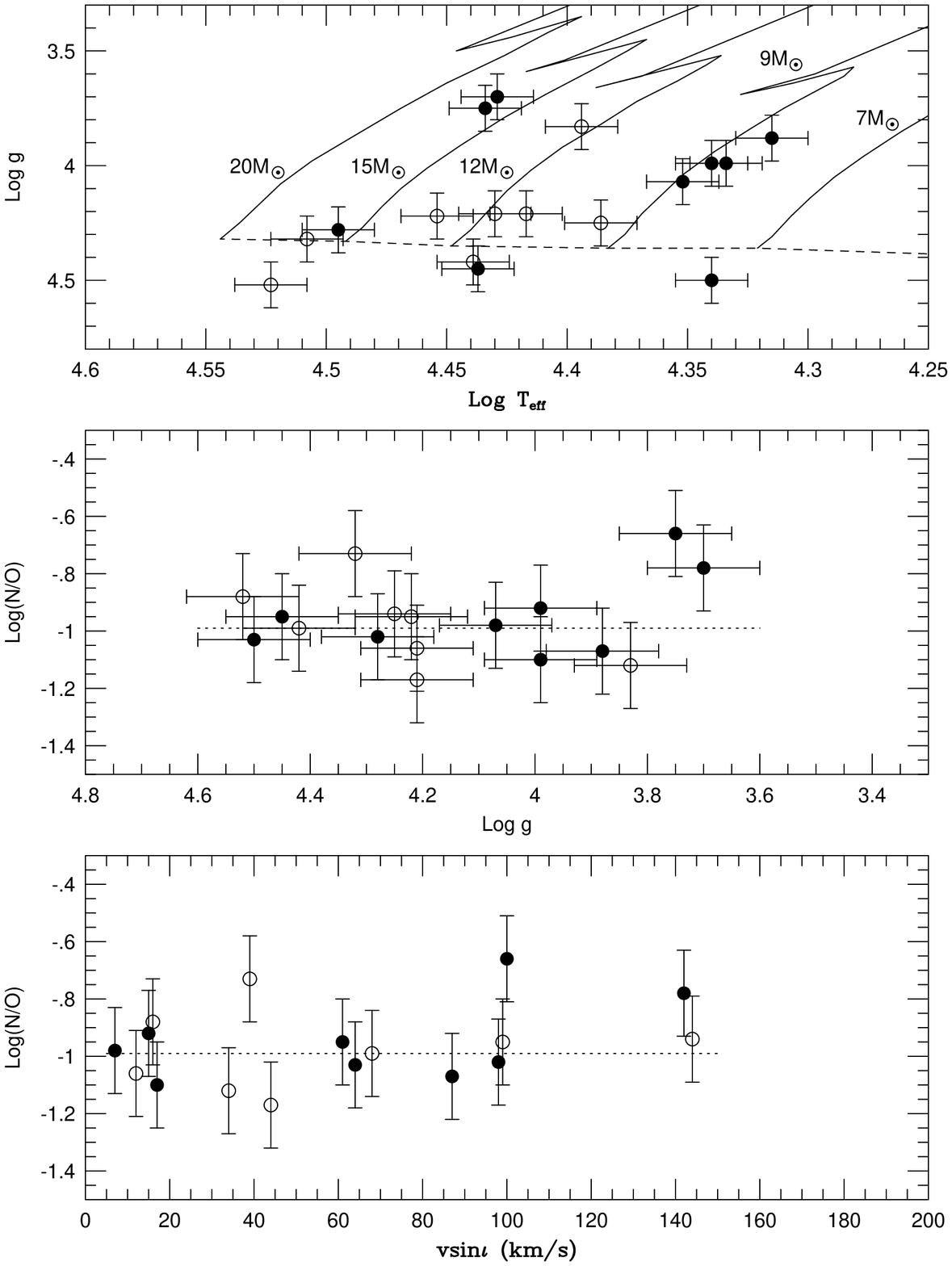,width=16cm,height=22cm} }
\caption{ }
\end{figure}

\end{document}